\documentclass{PoS}
\usepackage{amsfonts}
\usepackage{amsmath}
\usepackage{amssymb}

\newcommand{\boldsigma}{\mbox{\boldmath $\sigma$}}
\newcommand{\boldtau}{\mbox{\boldmath $\tau$}}
\newcommand{\boldpi}{\mbox{\boldmath $\pi$}}

\title{Lattice QCD and nuclear physics for searches of physics beyond the Standard Model}

\ShortTitle{Lattice QCD and nuclear physics for BSM searches}

\author{\speaker{Emanuele Mereghetti}\thanks{Work supported by the DOE Office of Science and by the LDRD program at Los Alamos National
Laboratory. }\\
        Los Alamos National Laboratory\\
        E-mail: \email{emereghetti@lanl.gov}}

\abstract{Low-energy tests of fundamental symmetries are extremely sensitive probes of physics beyond the Standard Model, reaching scales that are comparable, if not higher, than directly accessible at the energy frontier. The interpretation of low-energy precision experiments and their connection with models of physics beyond the Standard Model relies on controlling the theoretical uncertainties induced by the nonperturbative nature of QCD at low energy and of the nuclear interactions.

In these proceedings, I will discuss how the interplay of Lattice QCD and nuclear Effective Field Theories can lead to improved predictions for low-energy experiments, with controlled uncertainties. I will describe the framework of chiral Effective Field Theory, and then discuss a few examples, including non-standard $\beta$ decays, neutrinoless double beta decay and searches for electric dipole moments, to highlight the progress achieved in recent years, and 
the role that Lattice QCD will play in addressing the remaining open problems.}

\FullConference{The 36th Annual International Symposium on Lattice Field Theory - LATTICE2018\\
		22-28 July, 2018\\
		Michigan State University, East Lansing, Michigan, USA.}

\begin{document}

\section{Introduction}

The Standard Model (SM) of Particle Physics is an extremely successful theory, which has passed a number of stringent experimental tests \cite{ALEPH:2005ab,Tanabashi:2018oca,Bona:2006ah,Charles:2015gya,Amhis:2016xyh}.
The last missing ingredient of the SM, the Higgs boson, has been discovered at the Large Hadron Collider (LHC) \cite{Aad:2012tfa,Chatrchyan:2012xdj}, 
and, within the still relatively large theoretical uncertainties, its properties appear to be SM-like \cite{Khachatryan:2016vau,ATLAS:2018doi}. 
In addition to looking for and finding the Higgs, the experiments at the LHC are carrying out an extensive program of searches for new particles.
A quick look at the ATLAS and CMS exotics pages \cite{ATLASex} shows the breadth of this program, and, at the same time, the fact that no conclusive evidence of the production of beyond-the-SM (BSM) particles has  emerged.
We know, on the other hand, that the SM is not the complete theory of Nature. 
The neutrino oscillation experiments of the last two decades \cite{Eguchi:2002dm,Ahmad:2001an} have definitively shown that neutrino have masses,  
which are not accommodated in the minimal version of the SM. 
In addition, the SM cannot successfully generate the observed matter-antimatter asymmetry in the Universe,
lacking, in particular, a strong enough source of CP violation \cite{Gavela:1993ts,Gavela:1994ds,Gavela:1994dt,Huet:1994jb}, and the SM does not have a viable dark matter candidate. 

The origin of neutrino masses, baryogenesis and the nature of dark matter are three of the most important open questions in particle physics.
In all three cases, low-energy precision experiments with nuclear targets play a fundamental role, as they are highly competitive and complementary to collider experiments.
Neutrinoless double beta decay  experiments  have the best chance of determining whether neutrinos are Majorana particles. 
Current experimental limits on the double-beta half-life of several nuclei are already very stringent
\cite{Gando:2012zm,Agostini:2013mzu,Albert:2014awa,Andringa:2015tza,KamLAND-Zen:2016pfg,Elliott:2016ble,Agostini:2017iyd,Aalseth:2017btx, Albert:2017owj,Alduino:2017ehq,Agostini:2018tnm, Azzolini:2018dyb}, and will improve by one or two orders of magnitude in the next generation of experiments. 
The new sources of CP violation needed for baryogenesis might manifest themselves in searches for electric dipole moments (EDMs) of the neutron, proton, diamagnetic or paramagnetic atoms and molecules \cite{Andreev:2018ayy,Afach:2015sja,Graner:2016ses,Bishof:2016uqx,Cairncross:2017fip,Chupp:2017rkp}, while  dark matter direct detection experiments are rushing towards the ``neutrino floor'' \cite{Aprile:2018dbl,Akerib:2016vxi,Cui:2017nnn}, their ultimate irreducible background.

The challenge of low-energy experiments is that they are sensitive to a variety of scales, from the TeV or multi-TeV scale at which new physics arises, 
all the way down to the scales of nuclear and atomic physics. In particular sizable theoretical uncertainties are introduced by the nonperturbative nature of QCD at long distances, and 
by the nonperturbative nature of the nuclear force. 
In these proceedings I will review how the interplay between Lattice QCD (LQCD) and nuclear Effective Field Theories (EFTs) has
allowed to make  progress towards the goal of controlled theoretical predictions of low-energy nuclear observables.
In Section \ref{chi} I will briefly review the tools of chiral EFT. I will then discuss applications to three examples. In Section \ref{DM} I will
discuss the hadronic and nuclear physics input for dark matter (DM) direct detection and non-standard $\beta$ decay experiments. In these cases, the most important operators 
are in the single nucleon sector, and, thanks to progress in both LQCD and chiral EFT, the matching between the theories at the  quark and hadronic levels is now under control.
In Section \ref{0nbb} I will discuss neutrinoless double beta decay ($0\nu\beta\beta$). The leading order (LO) transition operators involve, in this case, two nucleons. 
I will discuss the complications that ensue from this feature, and the role that LQCD will have in resolving them.
Finally, in Section \ref{EDM} I will review recent progress in first principle calculations of the nucleon EDM, and of CP-violating (CPV) pion-nucleon couplings.

\section{Chiral Effective Field Theory}\label{chi}

The derivation of a systematic connection between Quantum Chromodynamics (QCD),
the microscopic theory of the strong interactions, and the dynamics of few-nucleon systems has 
undergone considerable progress thanks to the development of chiral Effective Field Theory  \cite{Weinberg:1990rz,Weinberg:1991um,Ordonez:1993tn}.
The starting point of chiral EFT is the approximate $SU(2)_L \times SU(2)_R$ symmetry of the QCD Lagrangian. Chiral symmetry, and its spontaneous breaking 
to the $SU(2)_V$ subgroup, imply the existence of three pseudo-Goldstone bosons, the pions, whose  couplings, to themselves and to
matter, vanish at zero momentum. In the mesonic and single nucleon sector, this property guarantees a perturbative expansion of amplitudes in powers of $Q/\Lambda_\chi$,
where $Q$ is a momentum scale of order of the pion mass, and $\Lambda_\chi \sim 4 \pi F_\pi \sim 1$ GeV is the breakdown scale of the theory.

For systems with two or more nucleons
the energy scale  $Q^2/2m_N$ becomes relevant 
and scattering amplitudes do not have a homogeneous scaling in $Q$.  
Therefore, the perturbative expansion of  interactions   
does not guarantee a perturbative expansion of amplitudes \cite{Weinberg:1990rz,Weinberg:1991um}.
Indeed, the class of diagrams in which the intermediate state consists purely of propagating nucleons,
the  so-called ``reducible'' diagrams,
is enhanced by factors of $m_N/Q$ with respect to the counting rules of chiral EFT in the single nucleon sector.
On the other hand, loop  diagrams whose intermediate states contain interacting nucleons and pions --``irreducible''-- 
follow the power counting of chiral EFT in the single nucleon sector~\cite{Weinberg:1990rz,Weinberg:1991um}.
Reducible diagrams are then obtained by patching together irreducible diagrams with intermediate states consisting of free-nucleon propagators.
This is equivalent to solving the Lippman-Schwinger equation with a potential defined by the sum of irreducible diagrams. 
This setup is shared by other non-relativistic systems, such as the hydrogen atom or heavy quarkonia, described by the effective theories of NRQED \cite{Labelle:1996en}, potential NRQED \cite{Pineda:1998kn}, NRQCD \cite{Bodwin:1994jh} and potential NRQCD \cite{Brambilla:1999xf}.

Weinberg's original proposal was to construct the strong interaction potential (and external currents)  in a perturbative expansion in $Q/\Lambda_\chi$.
Potentials and currents contains several undetermined couplings, usually dubbed ``low-energy constants'' (LECs). In Weinberg's power counting, the scaling of these couplings,
and therefore which operators need to be kept at each order, is determined by naive dimensional analysis  (NDA) \cite{Manohar:1983md}, and the LECs
are subsequently fit to experimental data, e.g. nucleon-nucleon scattering data.
The derivation of chiral EFT interactions \cite{Piarulli:2016vel,Reinert:2017usi,Entem:2017gor,Lynn:2017fxg}
and of SM electromagnetic and weak currents \cite{Pastore:2008ui,Kolling:2009iq,Baroni:2015uza,Krebs:2016rqz}
has reached a very high level of sophistication, giving an excellent description of the properties of few-nucleon systems.

On the other hand, a peculiarity of the nuclear interaction and of chiral EFT is that the LO potential is singular, because of the presence of LO contact interactions 
and because of the $1/r^3$ dependence of the tensor component of the pion exchange potential.
These singular potentials lead to divergences in the solution of the Lippman-Schwinger equation, whose taming requires the promotion
of certain contact interactions, formally subleading in Weinberg's counting, to LO \cite{Kaplan:1996xu,Nogga:2005hy,Long:2011xw,Long:2012ve}. 
Notwithstanding the phenomenological success of Weinberg's counting, the derivation of a consistent power counting  for the strong interaction potential and for external currents, 
allowing the theory to be renormalized, is necessary for the interpretation of chiral EFT as a quantum field theory. The study of the regulator independence
of amplitudes and observables is particularly important for the chiral realization of BSM operators, when consistency problems 
cannot be simply resolved by including enough operatorial structures and fitting their LECs to data, but the chiral EFT power counting should be the guiding principle
in assessing the importance of various interactions.

\section{Dark matter direct detection and non-standard $\beta$ decay}\label{DM}

The first examples we consider are DM direct detection 
experiments and searches for non-standard charged-current interactions in $\beta$ decay. These processes can be described in a model independent way 
by writing down effective interactions between colorless probes, such as electrons, neutrinos and DM particles, and quarks and gluons, which respect the gauge symmetries of the SM. 
These operators are organized according to their canonical dimension, with operators of higher dimension suppressed by more and more powers of $\Lambda$, 
the high-energy scale at which new physics arises.

For both $\beta$ decays and DM direct  detection, the operators of lowest dimension involve quark bilinears. For example, 
if one considers only left-handed neutrinos,
the charged-current Lagrangian at scales $\sim 1$ GeV starts at dimension-six, and contains, in addition to the SM $V-A$ current, 
a $V+A$ component, and scalar, pseudoscalar and tensor interactions \cite{Lee:1956qn,Cirigliano:2012ab,Gonzalez-Alonso:2018omy}:
\begin{eqnarray}
\mathcal L_\beta = - \frac{4 G_F}{\sqrt{2}} V_{ud} & & \Bigg\{ (1 + \epsilon_L) \bar \nu_L \gamma^\mu e_L  \bar d_L \gamma_\mu  u_L 
+ \epsilon_R \bar \nu_L \gamma^\mu e_L  \bar d_R \gamma_\mu  u_R
+ \frac{1}{2} (\epsilon_S - \epsilon_P) \bar \nu_L e_R \,  \bar d_R   u_L \nonumber  \\ & &  
+ \frac{1}{2} (\epsilon_S + \epsilon_P) \bar \nu_L e_R \,   \bar d_L   u_R +  \epsilon_T \bar \nu_L \sigma^{\mu \nu} e_R \,   \bar d_L  \sigma_{\mu \nu} u_R  \Bigg\}.
\label{eq:1}
\end{eqnarray}
The couplings $\epsilon_i$ contain information about new physics, and scale as $\epsilon_{i} \sim {v^2}/{\Lambda^2}$, where $v = 246$ GeV is the Higgs vacuum expectation value.
A very similar basis exists for \cite{Goodman:2010ku}, which includes, in addition, flavor singlet operators and the coupling of DM to gluons. 

For quark bilinear operators as those in Eq. \eqref{eq:1}, the most important 
hadronic effects are captured by one-body operators \cite{Cirigliano:2012ab,Cirigliano:2012pq,Cirigliano:2013xha,Hoferichter:2015ipa,Bishara:2016hek},
in particular the isoscalar and isovector vector, axial, scalar, pseudoscalar and tensor one-body currents.
In the case of the vector and axial currents, the single nucleon charges and form factors can be determined from experimental data,
with LQCD  becoming increasingly competitive  \cite{Chang:2018uxx,Gupta:2018qil,Rajan:2017lxk}. 
For the nucleon light-quark scalar charge,  there is a long-standing disagreement between the values obtained
in chiral perturbation theory or with dispersive techniques
\cite{Alarcon:2011zs,Hoferichter:2015dsa} and  LQCD \cite{Durr:2015dna,Yang:2015uis,Abdel-Rehim:2016won} (for a critical discussion, see Ref. \cite{Hoferichter:2016ocj}).
The resolution of this puzzle is crucial to reduce the theoretical uncertainties on the DM-nuclei scattering cross sections.
The determination of the tensor charges required  dedicated LQCD calculations \cite{Alexandrou:2017qyt,Gupta:2018qil,Gupta:2018lvp}, which have now reached  5\% precision for the 
$u$-  and $d$-quark charges.
Such control is necessary for low-energy precision experiments to compete with the strong constraints on non-standard charged-current interactions from the LHC 
\cite{Cirigliano:2012ab,Falkowski:2017pss,Gonzalez-Alonso:2018omy,Gupta:2018qil,Alioli:2018ljm}. 

While one-body operators tend to give the dominant contributions, two-body currents are important to reach agreement between theory and data 
for standard $\beta$ decays of light nuclei \cite{Pastore:2017uwc}. 
Two-body scalar currents have been studied in the context of DM scattering  \cite{Hoferichter:2016nvd,Hoferichter:2017olk,Korber:2017ery},
and found to be negligible in the case of light nuclei, and more important for heavy nuclei such as $^{132}$Xe. 
Finally, to our knowledge, there has not been a systematic study of two-body effects in tensor currents.
It is interesting to notice that the first calculations of the \textit{nuclear} axial, scalar and tensor charges on the lattice have appeared  \cite{Chang:2017eiq},
at  pion mass of $m_\pi = 806$ MeV.
As the pion mass will move towards the physical point, these calculations will
allow to test the chiral EFT power counting in few-nucleon systems, and will
serve as input for the calculation of BSM processes in  nuclei.

\section{Neutrinoless double beta decay }\label{0nbb}
We next discuss neutrinoless double beta decay. $0\nu\beta\beta$ is the most sensitive laboratory probe of lepton number violation (LNV).
Its observation will prove that neutrinos are Majorana particles, with important consequences for the understanding of the origin of neutrino masses,
and for the viability of leptogenesis scenarios, in which the generation of the matter-antimatter asymmetry in the Universe proceeds in the lepton sector. 

Lepton number is an accidental symmetry of the SM, and, as for DM-quark interactions or non-standard $\beta$ decay, LNV at the electroweak scale can be parametrized in terms of 
gauge invariant effective operators. The operator of lowest dimension is the dimension-five Weinberg's operator \cite{Weinberg:1979sa},
which, after electroweak symmetry breaking, induces neutrino masses and mixings. In particular, the LNV parameter that enters $0\nu\beta\beta$ is the electron-neutrino Majorana
mass
\begin{equation}\label{eq:intro.0}
\mathcal L_{\Delta L = 2} = - \frac{m_{\beta\beta}}{2} \nu_{eL}^T C \, \nu_{eL} + \ldots,
\end{equation}
where $m_{\beta\beta} = \sum U_{e i}^2 m_{i}$, $m_i$ are the masses of the neutrino mass eigenstates, and $U_{ei}$ are elements of the 
Pontecorvo-Maki-Nakagawa-Sato (PMNS) matrix. $C = i \gamma_2 \gamma_0$ denotes the charge conjugation matrix. 
Because of the  $SU(2)_L$ invariance of the SM, $m_i \sim v^2/\Lambda$.
Eq. \eqref{eq:intro.0} is the first term in an expansion in $v/\Lambda$, which includes operators of odd dimension  \cite{Kobach:2016ami}.
The full set of dimension-seven LNV operators is known \cite{Lehman:2014jma}, while subsets of operators of dimension-nine and higher have also been considered
\cite{deGouvea:2007qla,Graesser:2016bpz}. The implications  for $0\nu\beta\beta$ have been discussed in Refs. \cite{Pas:1999fc,Pas:2000vn,Prezeau:2003xn,Cirigliano:2017djv,Cirigliano:2018yza,Graf:2018ozy}.

We start by discussing the ``standard mechanism'' for $0\nu\beta\beta$, i.e. the exchange of light Majorana neutrinos.
Differently from Section \ref{DM}, in this case the transition is mediated by a two-body operator, which turns two neutrons in the initial state 
into two protons in the final state,
with the emission of two electrons. 
The $0\nu\beta\beta$ transition operator, or ``neutrino potential'', receives a long-range contribution from the exchange of 
off-shell Majorana neutrinos, with momentum $q \sim (0, k_F)$, where $k_F$ denotes the Fermi momentum $\sim 100$ MeV. 
The derivation of the long-range component of the neutrino potential has been known for a while 
\cite{Doi:1982dn,Haxton:1985am}. The standard result can be understood as the lowest order term in chiral EFT \cite{Cirigliano:2017tvr}.
Expressing the effective LNV Hamiltonian as 
\begin{eqnarray}
H_{\rm eff} =    2 G_F^2  V_{ud}^2 \  m_{\beta \beta}  
\  \bar e_{L} C \bar e_{L}^T   \ V_\nu  \,,
\label{eq:HV}
\end{eqnarray}
where $G_F$ is the Fermi constant and $V_{ud}$ the $ud$ element of the CKM matrix, 
the long- and pion-range contributions to the neutrino potential are given by
\begin{equation}
    V_{\nu}^{}  =   \tau^{(1)+} \tau^{(2)+}  
\,  \frac{1}{\vec{q}^2}  \,  \left\{
%g_V^2 -
1-  \frac{2}{3} g_A^2  \boldsigma^{(a)} \cdot \boldsigma^{(b)} \left( 1  +  \frac{1}{2}\frac{m_\pi^4}{ (\vec q^2 + m_\pi^2)^2} \right)  - 
\frac{1}{3} g_A^2 S^{(12)} \left( 1 - \frac{m_\pi^4}{ (\vec q^2 + m_\pi^2)^2}\right)     \right\}~.
 \label{eq:Vnu0}
\end{equation}
Here  $g_A \sim 1.27$ is the nucleon axial coupling, $m_\pi$ the pion mass and $S^{(12)} = - (3 \boldsigma^{(1)} \cdot \hat{\vec q} \, \boldsigma^{(2)} \cdot \hat{\vec q} - \boldsigma^{(1)} \cdot \boldsigma^{(2)})$.

Eq. \eqref{eq:Vnu0} receives three kind of corrections.
One can first of all consider corrections to the single nucleon axial and vector currents. These are usually included in the calculation
of nuclear matrix elements (NMEs) of the neutrino potential via axial and vector form factors, see for example \cite{Simkovic:1999re,Engel:2016xgb}.
There are then two-body corrections to the axial and vector currents, which give rise to three-body neutrino potentials \cite{Menendez:2011qq,Wang:2018htk}.
Finally there are ``non-factorizable'' two-body contributions, arising from pion-neutrino loops, and, intimately related to them, the contributions from ultrasoft neutrinos \cite{Cirigliano:2017tvr}.
The pion-neutrino loops are in general ultraviolet divergent, 
denoting sensitivity to short-range physics, and the divergences are canceled by local LNV  $\pi\pi e e$, $n p \pi e e$  and $n n pp e e$ operators.
The counterterm Lagrangian has the form 
\begin{equation}\label{eq:ct}
\mathcal L_{\rm ct} = 
\frac{2 G_F^2 V_{ud}^2  m_{\beta \beta}}{(4 \pi F_\pi)^2}
\left[    \frac{5}{6}  F_0^2   g_\nu^{\pi \pi} \,  \partial_\mu \pi^- \partial^\mu \pi^- 
+   \sqrt{2}g_A^0 F_0    g_\nu^{\pi N}  \   \bar p S_\mu n   \,  \partial^\mu  \pi^- 
+ g_\nu^{NN}   \bar p n \,  \bar{p} n   \right]  \,  \bar e_L C\bar e_L^T+\dots  \qquad 
\end{equation}
where $\ldots$ denote terms with additional pion fields, which are fixed by chiral symmetry, and $F_\pi \sim 92$ MeV is the pion decay constant.  The LECs $g_{\nu}^{\pi\pi}$, $g_{\nu}^{\pi N}$
and $g_{\nu}^{NN}$ are unknown, and need to be determined by LQCD calculations of LNV processes, such as $\pi^- \pi^- \rightarrow e^- e^-$ \cite{Feng:2018pdq}. 
In Weinberg's counting, $g_{\nu}^{\pi\pi}$, $g_{\nu}^{\pi N}$
and $g_{\nu}^{NN}$ should be $\mathcal O(1)$,  and affect the $0\nu\beta\beta$ half-life only at N$^2$LO.

Weinberg's power counting, however,  leads to inconsistencies  due to a conflict between NDA and the nonperturbative renormalization
required by the short-range core of the nuclear force. As shown in Ref. \cite{Kaplan:1996xu}, the problem is particularly serious for two nucleons in the $^1S_0$ channel,
the channel most relevant to $0\nu\beta\beta$. 
One can study the consistency of Weinberg's  counting for LNV matrix elements by examining the  $n n \rightarrow p p e^- e^-$
scattering process. We consider the LO strong interaction potential, which is particular simple in the $^1S_0$ channel,
\begin{equation}\label{eq:LNV2}
V_0(\vec p, \vec p^\prime) = \tilde{C}  - \frac{4\pi \alpha_\pi}{\vec q^2 + m_\pi^2}, \qquad \alpha_\pi = \frac{g_A^2 m_\pi^2}{16 \pi F_\pi^2}.
\end{equation}
The delta function potential in Eq. \eqref{eq:LNV2} can be regulated in a variety of ways, such as dimensional regularization  \cite{Kaplan:1996xu},
or with a Gaussian cut-off 
\begin{equation}
\delta^{(3)}_{R_S}(\vec r) =  \frac{1}{(\sqrt{\pi} R_S)^3} \exp\left(- \frac{r^2}{R_S^2}\right),
\end{equation}
and the LEC $\tilde C$ is determined by reproducing the $np$ scattering length in the $^1S_0$ channel, in a given regularization scheme.
The LNV amplitude is then defined as
\begin{eqnarray}\label{eq:anutot}
\mathcal A_\nu = -  \int d^3 \vec r \, \psi^-_{\vec p^\prime}(\vec r)^* \  V_\nu(\vec r)  \ \psi^+_{\vec p}(\vec r),
\end{eqnarray}
where $\psi^\pm$ are the solutions of the Schr\"odinger equation with the potential in Eq. \eqref{eq:LNV2} which respect the correct  in- and out- asymptotic boundary 
conditions.

\begin{figure}[t]
\center
\includegraphics[width=10cm]{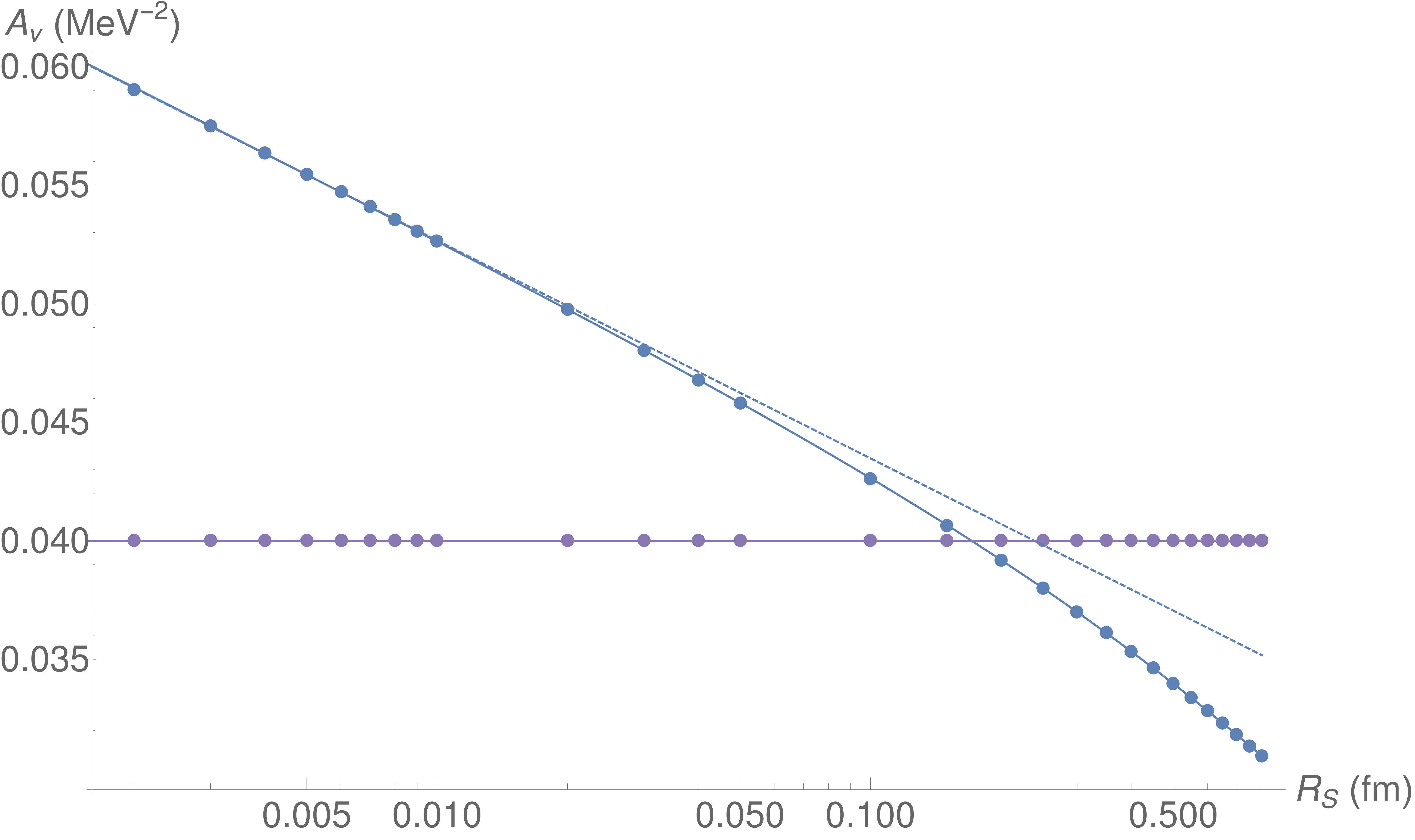}
\caption{Matrix element of the $0\nu\beta\beta$ transition operator $V_\nu$, defined in Eq. \eqref{eq:Vnu0},
for $\vec p = 1$ MeV and $\vec p^\prime = 38$ MeV.
The blue dots denote the matrix element computed using Eq. \eqref{eq:anutot}.
The dashed line is a fit to the functional form $ a + b \log R_S$, while the solid blue line includes additional power-suppressed terms \cite{Cirigliano:2018hja}. 
The purple dots and line show that the amplitude can be made cut-off independent by including the counterterm $g_{\nu}^{NN}$ at LO.
$g_{\nu}^{NN}$ was fixed by requiring $\mathcal A_\nu$ to be equal to $0.04$ MeV$^2$ at $\vec p = 1$ MeV, $\vec p^\prime = 38$ MeV.
}\label{Fig.1}
\end{figure}

The dependence of $\mathcal A_\nu$ on the cut-off $R_S$ is shown by the blue dots in Fig. \ref{Fig.1}. Fig. \ref{Fig.1} clearly shows  that, even after the strong scattering amplitude
is properly renormalized, the matrix element of the long-range neutrino potential is still logarithmically divergent. The logarithmic divergence can be canceled by promoting
$g_{\nu}^{NN}$ to leading order, that is, $g_{\nu}^{NN} \sim \mathcal O((4\pi)^2)$ rather than $\mathcal O(1)$. The  purple dots in Fig. \ref{Fig.1}
indeed show that, after the inclusion of $g_{\nu}^{NN}$, the amplitude can be made $R_S$-independent. 
This conclusion holds both in pionless and pionful EFTs 
\cite{Cirigliano:2017tvr,Cirigliano:2018hja}, and it is independent of the regulator used.

The study of the consistent renormalization of LNV amplitudes in chiral EFT thus points out 
that the $0\nu\beta\beta$ transition operator has a LO short-range component, which 
has not been included in all existing calculations of $0\nu\beta\beta$ NMEs. While the renormalization group equation of $g_{\nu}^{NN}$ can be derived \cite{Cirigliano:2017tvr,Cirigliano:2018hja}, 
the determination of the finite part, and thus a solid assessment of the numerical impact of $g_{\nu}^{NN}$ on the half-life of nuclei of experimental interest,
requires the calculation of the $n n \rightarrow p p e^- e^-$ scattering amplitude on the lattice, to be then matched   onto chiral EFT. 
While these are very difficult calculations, the LQCD community is already starting to attack them. For example, a preliminary calculation of the process $\pi\pi \rightarrow e e$
recently appeared \cite{Feng:2018pdq}, allowing the determination of $g_{\nu}^{\pi\pi}$ in Eq. \eqref{eq:ct},
and the NPLQCD collaboration has already studied processes with insertions of two weak currents, which
are important for two-neutrino double beta decay \cite{Shanahan:2017bgi,Tiburzi:2017iux}.

The light Majorana neutrino exchange mechanism dominates if the mass scale at which LNV arises 
is very high, $\Lambda \sim 10^{14}$ GeV. 
In many models, the scale of LNV can be much lower, and, in addition, the dimension-five Weinberg operator can be suppressed by small Yukawa couplings \cite{Tello:2010am,Nemevsek:2012iq}, 
or by loop factors \cite{Zee:1985id,Zee:1980ai,Babu:1988ki,Babu:1988ig,Babu:1988wk}. In these models 
the $0\nu\beta\beta$ half-life might receive contributions from dimension-seven and dimension-nine operators that are competitive or dominant  
with respect to the Weinberg operator. It is then important to classify the contributions of higher-dimensional operators 
and carefully identify the hadronic input required for a reliable estimate of $0\nu\beta\beta$ NMEs.

This task can once again  be accomplished using the tools of EFTs \cite{Prezeau:2003xn,Cirigliano:2017djv,Cirigliano:2018yza}.
Dimension-seven operators mainly require the hadronization of quark bilinears \cite{Cirigliano:2017djv,Cirigliano:2018yza},
which, as discussed in Section \ref{DM}, is under control.  On the other hand, dimension-nine operators 
involve four-quark operators, that match onto low-energy $\pi \pi e e$, $ n p \pi e e$ and $nn pp  ee$ interactions.
The calculation of the LECs  once again requires LQCD. The $\pi\pi$ matrix elements have been computed by the CalLat collaboration \cite{Nicholson:2018mwc},
with a few-percent accuracy. This calculation agrees with the less precise estimates of \cite{Cirigliano:2017ymo,Savage:1998yh}, based on 
$SU(3)$ relations between the four-quark operators for $0\nu\beta\beta$, $\Delta S=2$ operators  that induce BSM contributions to $K$-$\bar K$ oscillations,  
and $\Delta S=1$ operators that contribute to $K \rightarrow \pi \pi$.

In Weinberg's power counting, the $\pi\pi e  e$ operators induce, in most cases, the dominant contribution to the NMEs \cite{Prezeau:2003xn}.
Pion-range transition operators are, however, affected by the same breaking of Weinberg's counting discussed for the light neutrino exchange mechanism \cite{Cirigliano:2018yza},
so that $nn pp ee$ operators cannot be neglected. 

This brief discussion showed how the interplay between LQCD, chiral EFT and many body methods is necessary for
reducing the large theoretical uncertainties that affect predictions of the  $0\nu\beta\beta$ half-life,
a crucial step to be able to extract fundamental LNV parameters, such as $m_{\beta\beta}$, from $0\nu\beta\beta$ experiments.

\section{Electric dipole moments}\label{EDM}

Electric dipole moments (EDMs) are sensitive probes of CP violation beyond the SM. 
EDMs experiments have put astounding limits on the EDMs of the electron, $d_e < 1.1 \cdot 10^{-16}$ $e$ fm \cite{Andreev:2018ayy},
neutron, $d_n < 3.0 \cdot 10^{-13}$ $e$ fm \cite{Afach:2015sja}, and of diamagnetic systems as $^{199}$Hg, $d_{^{199}\rm Hg} < 6.2 \cdot 10^{-17}$ $e$ fm
\cite{Graner:2016ses}, and $^{225}$Ra,  $d_{^{225}\rm Ra} < 1.2 \cdot 10^{-10}$ $e$ fm  \cite{Bishof:2016uqx}.  These limits are still six/seven orders of magnitude away from the SM 
predictions \cite{Pospelov:2005pr,Seng:2014lea}, leaving a large window for new physics. 

While the observation of an EDM in any of the aforementioned systems will unambiguously indicate the existence of CPV  beyond the SM, 
the EDM constraints  on specific new physics models are weakened by the large theoretical uncertainties that affect  the matrix elements of CPV  operators
between hadronic states. For example, EDM constraints on CPV in the top and Higgs sectors can vary by a factor of ten or more depending on whether 
theoretical uncertainties are considered \cite{Cirigliano:2016nyn,Chien:2015xha}.

CPV effects at the electroweak scale are captured by  $SU(3)_c \times SU(2)_L \times U(1)_Y$-invariant operators, 
which can then be matched  onto  an $SU(3)_c \times U(1)_{\rm em}$-invariant EFT \cite{deVries:2012ab}
( for a similar setup in $B$ physics, see Refs. \cite{Aebischer:2015fzz,Aebischer:2017gaw}, and for the fully general matching of the SM-EFT onto the low-energy 
$SU(3)_c \times U(1)_{\rm em}$-invariant EFT see Refs. \cite{Jenkins:2017jig,Jenkins:2017dyc}).
The QCD Lagrangian contains a single dimension-four operator, the QCD $\bar\theta$ term   \cite{Callan:1976je,tHooft:1976rip,tHooft:1976snw}, 
which can be rotated into a complex mass term, giving \cite{Baluni:1978rf}
\begin{eqnarray}
\mathcal L_{4} = m_* \bar\theta\, \bar q i \gamma_5 q,  \qquad
m_* = \frac{m_u m_d m_s}{ m_s (m_u + m_d) + m_u m_d} = \frac{\bar m (1 - \epsilon^2)}{2  + \frac{\bar m}{m_s} (1-\epsilon^2)}.
\end{eqnarray}
The combinations of light quarks masses $\bar m$ and $\epsilon$ are $2 \bar m = m_u + m_d$, $\epsilon = (m_d - m_u)/(m_d + m_u)$. 
Focusing on the purely hadronic sector, 
the low-energy operators that are induced by SM-EFT operators at tree level are \cite{deVries:2012ab,Jenkins:2017jig,Mereghetti:2018oxv}
\begin{eqnarray}\label{quark}
\mathcal L_{6} &=&  \frac{g_s C_{\tilde{G}}}{6 v^2} f^{a b c} \epsilon^{\mu \nu \alpha \beta}  G^a_{\alpha \beta} G_{\mu \rho}^{b} G^{c\, \rho}_{\nu}   
\nonumber \\ & & -\sum_{q}\frac{ m_q }{2 v^2} \left( \tilde c^{(q)}_\gamma \bar{q} i \sigma^{\mu\nu} \gamma_5 q \; e F_{\mu\nu}  
+ \tilde c^{(q)}_g \bar{q} i \sigma^{\mu\nu} \; g_s G_{\mu\nu} \gamma_5  q \right) \nonumber\\
& & 
- \frac{4 G_F}{\sqrt{2}} \Bigg\{ \Sigma^{(ud)}_1 (\bar d_L u_R \bar u_L d_R - \bar u_L u_R \bar d_L d_R )   +    \Sigma^{(us)}_1   (\bar s_L u_R \bar u_L s_R - \bar s_L s_R \bar u_L u_R)   \nonumber \\
& & + \Sigma^{(ud)}_2 (\bar d^\alpha_L u^\beta_R\, \bar u^\beta_L d^\alpha_R - \bar u^\alpha_L u^\beta_R\, \bar d^\beta_L d^\alpha_R )   +    \Sigma^{(us)}_2   (\bar s^\alpha_L u^\beta_R \, \bar u^\beta_L s^\alpha_R 
- \bar s^\alpha_L s^\beta_R \, \bar u^\beta_L u^\alpha_R) \nonumber \\
& & +    \Sigma^{(us)}_3   (\bar s_L u_R \bar u_L s_R + \bar s_L s_R \bar u_L u_R) +     \Sigma^{(us)}_4   (\bar s^\alpha_L u^\beta_R \, \bar u^\beta_L s^\alpha_R + \bar s^\alpha_L s^\beta_R \, \bar u^\beta_L u^\alpha_R) 
\Bigg\} \nonumber \\
& &- \frac{4 G_F}{\sqrt{2}} \Bigg\{  \Xi^{(ud)}_1 \, \bar d_L \gamma^\mu  u_L\, \bar u_R \gamma_\mu  d_R    + \Xi^{(us)}_1  \bar s_L \gamma^\mu  u_L\, \bar u_R \gamma_\mu  s_R  
+ \Xi^{(ds)}_1  \bar s_L \gamma^\mu  d_L\, \bar d_R \gamma_\mu  s_R  \nonumber \\
& & 
\Xi^{(ud)}_2 \, \bar d^\alpha_L \gamma^\mu  u^\beta_L\, \bar u^\beta_R \gamma_\mu  d^\alpha_R    + \Xi^{(us)}_2  \bar s^\alpha_L \gamma^\mu  u^\beta_L\, \bar u^\beta_R \gamma_\mu  s_R^\alpha  
+ \Xi^{(ds)}_2  \bar s^\alpha_L \gamma^\mu  d^\beta_L\, \bar d^\beta_R \gamma_\mu  s^\alpha_R
\Bigg\}.
\end{eqnarray}
Eq. \eqref{quark} includes the quark EDM (qEDM) and chromo-EDM (qCEDM) operators , $\tilde c^{(q)}_{\gamma, g}$,
the Weinberg three-gluon operator, $C_{\tilde G}$, six $LR\,LR$ and six $LL\, RR$ four-quark operators.
The coefficients $\tilde C_G$, $\tilde c^{(q)}_{\gamma, g}$, $\Sigma^{(q q^\prime)}_{1,2,3,4}$ and $\Xi^{(q q^\prime)}_{1,2}$ are dimensionless, and scale as $(v/\Lambda)^2$.

The operators in Eq. \eqref{quark} need to be matched onto chiral EFT. In particular, 
the chiral Lagrangian relevant for the calculation of the LO EDMs of the nucleon and of light nuclei is \cite{deVries:2012ab}
\begin{eqnarray}\label{chiral}
\mathcal L_{\chi}= - 2 \bar N \left( \bar{d}_0  + \bar d_1 \tau_3 \right) S^{\mu} v^{\nu} N F_{\mu \nu}  - \frac{\bar g_0}{2 F_{\pi}} \bar N \boldpi \cdot \boldtau N - \frac{\bar g_1}{2 F_{\pi}} \pi_3 \bar N N,  
\end{eqnarray}
where $N$ denotes the nucleon doublet $N = (p, n)^T$. 
$\bar d_{0,1}$ denote short-range contributions to the nucleon EDM, while $\bar g_{0,1}$ are CPV pion-nucleon couplings, which give long-range contributions to the nucleon EDM \cite{Crewther:1979pi}
and to the CPV nucleon-nucleon potential \cite{Maekawa:2011vs,deVries:2012ab}.

The LECs in Eq. \eqref{chiral} depend on the source of CPV, and on the dynamics of the strong interactions. The transformation properties of 
the CPV  sources under chiral symmetry and isospin imply different hierarchies between the couplings in Eq. \eqref{chiral},
and predict different relations between EDMs in the one, two and three nucleon systems. The observation of such hierarchies in EDM experiments would then allow to disentangle 
quark-level operators,  pointing to different microscopic origin of CPV.

The feasibility of this program relies on the precise determination of the nucleon and nuclear EDMs as a function of the quark-level couplings in Eq. \eqref{quark},
going beyond the NDA estimates of Ref. \cite{deVries:2012ab}.
The calculation of EDMs of the nucleon and of light nuclei in terms of the couplings in Eq. \eqref{chiral} has reached a very satisfactory level of accuracy 
in recent years, thanks to a concerted effort in the chiral EFT community 
\cite{Liu:2004tq,Stetcu:2008vt,deVries:2011re,Bsaisou:2012rg,deVries:2011an,Bsaisou:2014zwa,Bsaisou:2014oka,Yamanaka:2015qfa,Yamanaka:2015ncb,Yamanaka:2016itb}.

The status of the determination of the LECs in Eq. \eqref{chiral} as a function of the parameters in Eq. \eqref{quark} is less satisfactory.
The only operators for which the nucleon EDM is known well are the qEDM operators $\tilde c_\gamma^{(u,d)}$. In this case, the relevant nucleon matrix element 
is the nucleon tensor charge, which, for the $u$ and $d$ quarks is known at the $5\%$ level \cite{Alexandrou:2017qyt,Gupta:2018lvp,Gupta:2018qil}.
The error on the contribution of $\tilde c_\gamma^{(s)}$ is larger, but both Ref. \cite{Alexandrou:2017qyt} and \cite{Gupta:2018qil} observe a non-zero signal.

The LQCD community has invested a considerable amount of effort to compute the nucleon EDM induced by the QCD $\bar\theta$ term \cite{Shintani:2005xg,Shintani:2006xr,Shintani:2008nt,Shintani:2015vsx,Shindler:2015aqa,Dragos:2017wms,Guo:2015tla,Abramczyk:2017oxr,Dragos:2018uzd}. 
At the moment, however, all calculations give results compatible with zero, as discussed in Ref. \cite{Abramczyk:2017oxr}.
The study of the nucleon EDM induced by $\tilde c^{(q)}_g$ and $\tilde C_G$ is also a very active research area \cite{Abramczyk:2017oxr,Izubuchi:2017evl,Shindler:2014oha,Bhattacharya:2016rrc,Rizik:2018lrz,Kim:2018rce}.  
Since the LQCD results are not yet conclusive, the best estimates remain  those derived with QCD sum rules  \cite{Pospelov:2005pr,Pospelov:2000bw,Demir:2002gg},
which however have large, $\mathcal O(100\%)$, uncertainties.

Chiral symmetry simplifies the determination of CPV pion-nucleon couplings.
As noticed in Ref. \cite{Crewther:1979pi} in the case of the QCD $\bar\theta$ term,  pion-nucleon couplings are related to the baryon spectrum.
For example, for the QCD $\bar\theta$ term  one can prove that  
\begin{equation}\label{g0theta}
\frac{\bar g_0}{2 F_\pi}(\bar\theta) = \frac{(m_n - m_p)_{\rm str}}{2 F_\pi} \frac{1-\epsilon^2}{2\epsilon}\, \bar\theta,
\end{equation}
where $(m_n - m_p)_{\rm str}$ is the contribution to the nucleon mass difference induced by $m_d - m_u$.
Eq. \eqref{g0theta} holds in both $SU(2)$ and $SU(3)$ chiral perturbation theory \cite{deVries:2015una}, up to small N$^2$LO corrections.
Extracting 
$(m_n - m_p)_{\rm str}$ from existing LQCD calculations
\cite{Borsanyi:2013lga,Borsanyi:2014jba,Brantley:2016our} yields a precise value for $\bar g_0$:
 \begin{equation}
\frac{\bar g_0}{2 F_\pi}(\bar\theta) = (15.5 \pm 2.0 \pm 1.6 ) \cdot 10^{-3}\, \bar\theta,
\end{equation}
where the first error is the LQCD error on $(m_n - m_p)_{\rm str}$, while the second is a conservative estimate of the error due to missing N$^{2}$LO terms in chiral perturbation theory.

Similarly, the CPV couplings induced by $\tilde c^{(q)}_{g}$, $\Xi^{(ij)}_{1,2}$  and $\Sigma^{(us)}_{3,4}$ can be extracted from modifications to the baryon spectrum
induced by the CP-conserving chiral partners of these operators \cite{deVries:2016jox,Cirigliano:2016yhc,Seng:2016pfd}.
For example, in the case of the qCEDM and of the $LL\, RR$ operators $\Xi$ 
one finds that $\bar g_{0,1}$ are given by  \cite{deVries:2016jox,Cirigliano:2016yhc}
\begin{eqnarray}\label{eq:g0g1_mod}
\bar g_0 &=& 
\Bigg\{ \tilde d_0\left( \frac{d }{d  c_3} + r \frac{d}{d (\bar m \varepsilon)}\right)
- \sum_{i=1,2} \textrm{Im}\, \Xi^{(us)}_{i}   \left( \frac{d}{d \textrm{Re}\, \Xi^{(us)}_{i}} + \frac{r_i}{2 v^2} \frac{d}{d \bar m \varepsilon} \right)
\nonumber \\ & & - \sum_{i=1,2} \textrm{Im}\, \Xi^{(ds)}_{i}   \left( \frac{d}{d \textrm{Re}\, \Xi^{(ds)}_{i}} + \frac{r_i}{2 v^2} \frac{d}{d \bar m \varepsilon} \right)
\Bigg\}  (m_n - m_p)  + \delta m_{N,\textrm{QCD}} \frac{1-\varepsilon^2}{2\varepsilon} \left(\bar\theta-\bar\theta_{\mathrm{ind}}\right)\ ,\nonumber \\
\bar g_1 &=& \Bigg\{ - \tilde d_3 \left( \frac{d}{d  c_0} - r \frac{d}{d \bar m} \right)
+ \sum_{i=1,2}\textrm{Im}\, \Xi^{(us)}_{i}   \left( \frac{d}{d \textrm{Re}\, \Xi^{(us)}_{i}} - \frac{r_i}{2 v^2} \frac{d}{d \bar m } \right) \nonumber \\
& & 
- \sum_{i=1,2}\textrm{Im}\, \Xi^{(ds)}_{i}   \left( \frac{d}{d \textrm{Re}\, \Xi^{(ds)}_{i}} - \frac{r_i}{2 v^2} \frac{d}{d \bar m } \right)
+ 2  \sum_{i=1,2}\textrm{Im}\, \Xi^{(ud)}_{i}   \left( \frac{d}{d \textrm{Re}\, \Xi^{ud}_{i}} - \frac{r_i}{2 v^2} \frac{d}{d \bar m } \right)
\Bigg\} (m_n + m_p)\ . \nonumber \\
\end{eqnarray}
In Eq. \eqref{eq:g0g1_mod}  we introduced 
the CP-even chiral partners of the qCEDM 
\begin{equation}
\mathcal L = -\sum_{q}\frac{ m_q }{2 v^2}\,   c^{(q)}_g \bar{q} i \sigma^{\mu\nu} \; g_s G_{\mu\nu}  q 
\end{equation}
and defined the couplings $v^2 \tilde d_{0,3} = m_u \tilde c_g^{(u)} \pm m_d \tilde c_g^{(d)}$,  and $v^2 c_{0,3} = m_u  c_g^{(u)} \pm m_d  c_g^{(d)}$.  
$r$, $r_{1,2}$ are ratios of vacuum matrix elements of the qCEDM and $LL\,RR$ operators, which can also be expressed in terms of derivatives of the pion mass. For example, for the qCEDM, 
\begin{equation}\label{eq:1.2}
r = \frac{1}{2} \frac{ \langle 0 | \bar q  g_s \sigma_{\mu \nu}   \, G^{\mu \nu} q  | 0 \rangle}{\langle 0 | \bar q q | 0 \rangle}  = \frac{d m_\pi^2}{d c_0 } \frac{d \bar m}{d m_\pi^2}.
\end{equation}
Finally, $\bar\theta_{\mathrm{ind}}$ is the minimum of the axion potential in the presence of dimension-six operators
\begin{eqnarray}\label{eq:1.3}
\bar\theta_{\mathrm{ind}} &=&   \frac{r}{v^2} \left( \tilde c^{(u)}_g + \tilde c^{(d)}_g + \tilde c^{(s)}_g\right) - \sum_{i=1,2} \frac{2 r_i}{v^2} \textrm{Im}
\Bigg[
\left( \frac{m_d - m_u}{2 m_u m_d}\right) \Xi^{(ud)}_{i} +
\left( \frac{m_s - m_u}{2 m_u m_s}\right) \Xi^{(us)}_{i} \nonumber \\ & & +
\left( \frac{m_s - m_d}{2 m_d m_s}\right) \Xi^{(ds)}_{i}
\Bigg]\ .
\end{eqnarray}
If the Peccei-Quinn mechanism is active \cite{Peccei:1977hh},  $\bar\theta$ relaxes to $\bar\theta_{\rm ind}$.
Eqs. \eqref{eq:g0g1_mod} and \eqref{eq:1.2} show that the CPV pion-nucleon couplings induced by chiral breaking operators are determined by the pion and nucleon 
generalized sigma terms. 
These generalized sigma terms should be  easily accessible on the lattice  \cite{deVries:2016jox},
providing a viable path for a  systematically improvable determination of $\bar g_{0,1}$.

\section{Conclusion}

Low-energy tests of fundamental symmetries are extremely sensitive probes of physics beyond the Standard Model, reaching scales that are comparable, if not higher, than directly accessible at 
collider experiments such as the LHC. The interpretation of low-energy  experiments and their 
implications for BSM physics relies on controlling the theoretical uncertainties induced by nonperturbative QCD and nuclear physics.
In these proceedings, I have discussed how the interplay between LQCD and nuclear EFTs can play a crucial role in bringing the theoretical 
uncertainties that affect low-energy observables, such as hadronic EDMs or the $0\nu\beta\beta$ half-life, under control.

\end{document}